\documentstyle[12pt]{article}
   \newcommand{\STE}{{\sin}^2\overline{\theta}_W}
   \newcommand{\STEE}{{\sin}^2\overline{\overline{\theta}}_W}
      \newcommand{\STW}{\sin^2\theta_W}
       \newcommand{\CTW}{\cos^2\theta_W}
      \newcommand{\SW}{\sin\theta_W}
     \newcommand{\SLT}{{S^\tau_L}^2}
      \newcommand{\SRT}{{S^\tau_R}^2}
    \newcommand{\SLB}{{S^b_L}^2}
    \newcommand{\SRB}{{S^b_R}^2}
\newcommand {\CT} {\cos \theta_2}
\newcommand {\ST} {\sin \theta_2}
\begin{document}
\vspace*{-1in}
\begin{flushright}
CERN-TH. 7196/94 \\
\end{flushright}
\vskip 90pt
\begin{center}
{\Large{\bf $Z^0$--${Z^0}^\prime$ and ordinary-exotic fermion
         mixings \\revisited using some special observables }}
\vskip 30pt
Gautam Bhattacharyya ${}^{*)}$
\vskip 10pt
{\it Theory Division, CERN, \\ CH 1211, Geneva 23, \\ SWITZERLAND}
\vskip 80pt

{\bf ABSTRACT}
\end{center}

We investigate the mixing of an extra ${Z^0}^\prime$
with the standard $Z^0$ and mixings of
 exotic fermions with their standard counterparts through some
  precisely measured electroweak
observables. These observables are geared to search for physics beyond
 the Standard
Model. We observe that although most of such mixings  are
severely constrained by the recent LEP data,
some of the mixing angles could still be rather large, awaiting
future tests.

\vskip 130pt
\begin{flushleft}
CERN-TH. 7196/94 \\
March 1994\\
\vskip 10pt
${}^{*)}$ gautam@cernvm.cern.ch
\end{flushleft}

\newpage
\setcounter{page} 1
 \pagestyle{plain}

\section{Introduction}
\par

 Precision measurements on and around the
 $Z$ peak at
 LEP have already put severe constraints on the
  possible mixing effects
  of
 the standard $Z^0$ with one extra neutral vector boson
  ${Z^0}^\prime$ and
 the mixings of sequential (ordinary) fermions with their exotic
  counterparts.
 These additional fermions and neutral bosons may stem
from many a theoretical extension  of the Standard Model (SM), the
most popular ones being the superstring-inspired $E_6$ \cite{u1}
 and the left--right symmetric (LRS) model \cite{lrm} which contain
the SM gauge group within their group structures.
 During each phase of the LEP run these analyses were carried
  out [3--12].
 Bounds have been put on $Z^0$--${Z^0}^\prime$
 mixing angles in refs. [3--6] and on ordinary-exotic mixing
angles in refs. [7--9] (on both in some refs.),
 by comparing the modified theoretical predictions of
 $ e^+e^- \rightarrow Z \rightarrow f\overline f$ cross sections
  and forward--backward asymmetries on and around
 the $Z$ peak with the corresponding experimental results.
  Variation of
  those
 bounds with $m_t, M_H$ and $\alpha_S$ have also been studied.
  The authors of
 refs. [10--12] have introduced a special set
  of observables, which are specifically geared to study these
   extensions beyond the (SM)
 with a rather focused attention. These new observables are
  constructed by
 combining the standard ones, which are directly probed by experiments,
mainly at LEP. One of the most interesting features of these
 observables
is that they can disentangle the two different sources of
 quadratic $m_t$
dependences in neutral current processes, appearing
through (i) $\Delta \rho \equiv \rho - 1$ (originating from
 two-point $W$
and $Z$ vacuum polarizations) and (ii) $\Delta V_b^t$
 (arising out of the
$Zb\overline b$ vertex correction). Refs. \cite{chiap1,chiap2} deal
basically with extensions motivated from $E_6$.
 In this paper we intend to impose bounds on more general kinds of
mixing effects, updating
existing bounds in some cases, using the latest LEP data \cite{lep}.
As tools for our studies we use some currently fashionable observables
which enjoy measurements of unprecedented precision at LEP.


\section{The observables and their SM expressions}
\par

Before getting down to the business of new physics
 we first recall the different LEP observables that interest us and
  their SM
expressions. The partial width of $Z$ to any particular
 standard fermion
flavour $i$ is given by
\begin{equation}
\Gamma_i = N^i_c {G_\mu M^3_Z \over 6\pi \sqrt {2}} (v_i^2 + a_i^2),
\end{equation}
 where
\begin{eqnarray}
 N^i_c &= & 1 + {3\alpha \over 4\pi} Q^2_i, ~~~~~~~~~~~~~~~~
{}~~~~~~~~~~~~~~~~~~(i= \mbox{\rm  lepton}),\nonumber \\
& = & 3\left(1 + {3\alpha \over 4\pi} Q_i^2\right)
\left[1 + {\alpha_S (M_Z)\over \pi} + ...\right],
 ~~~(i = \mbox{\rm  quark}).
\end{eqnarray}
The vector ($v_i$) and axial vector ($a_i$) couplings in the SM
 are given by
\begin{eqnarray}
{v_i}^{\mathrm {SM}} & = & \sqrt{\rho} \left[t_3^i -
 2 Q_i \STE \right] \nonumber \\
{a_i}^{\mathrm {SM}} & = & \sqrt{\rho}~ t_3^i
\label{smcoupl}
\end{eqnarray}
where $\sqrt\rho$ represents a non-trivial wave function
 renormalization of the
on-shell $Z$. In the framework of the SM $\rho$ is unity
at the tree level and it experiences a quantum correction
 ($\Delta \rho_t$) mainly due
 to the $t$--$b$ mass
splitting. To a very good approximation
\begin{equation}
\Delta\rho_t \simeq \frac{3 G_\mu m_t^2}{8 \pi^2 \sqrt{2}}.
\end{equation}
After improved Born approxmiation $\STE$ is
 different from $S_0^2$ ($\simeq 0.23$),
and for all practical purposes, $\STE \simeq
S_0^2 - \frac{3}{8} \Delta\rho_t$.
The couplings of $Z$ to $b$ quarks require a further
 correction, which
 arises from
 the $t$-mediated triangle loop in the above vertex. Since the $t$ quark
  couples in
full strength with the $b$ quark through $W$ exchange,
 this non-universal
correction is important only for the external $b$ lines. The modified
couplings are given by
\begin{eqnarray}
v_b & = & v_d - \frac{19}{60} \Delta V_b^t \nonumber \\
a_b  & = & a_d - \frac{19}{60} \Delta V_b^t,
\end{eqnarray}
where
\begin{equation}
\Delta V_b^t = -\frac{20 \alpha}{19 \pi}
\left( \frac{m^2_t}
{M^2_Z}
 + \frac{13}{6} \mbox{\rm  ln} \frac{m^2_t}{M^2_Z}\right).
\end{equation}

We now introduce four observables which are sensitive to many
 a different
 kind of new physics, including the ones we are probing for. We also
specify the virtues of these observables. The first
 one ($\widetilde{\gamma}_e$),
 which is called
the reduced electronic width \cite{chiap1}, is given by

\begin{equation}
\widetilde {\gamma}_e = \gamma_e - \frac {2}{3} \xi
\end{equation}
where
\begin{eqnarray}
\gamma_e &= &\frac{ 9~\Gamma_e}{\alpha(M_Z)~M_Z};~~~
\xi = \frac{M^2_W}{M^2_Z~ C^2_0};~~ C_0^2 = 1-S_0^2.
\end{eqnarray}
In the SM:
\begin{equation}
{\widetilde{\gamma}_e}^{\mathrm {SM}} =
\frac{1}{3} + \frac{2 v_0}{3}, ~~
\mbox{\rm  where}~~ v_0 = 1 - 4 S_0^2.
\end{equation}
Notice that $\widetilde{\gamma}_e$ is free from
 uncertainties originating
from $m_t$ or $M_H$.

The second one \cite{chiap1} $R_{\tau e} =
 \frac{\Gamma_\tau}{\Gamma_e}$
is a measure of any violation of lepton universality on the $Z$ peak.
In the SM (neglecting the $\tau$ mass), it is given by
\begin{equation}
{R_{\tau e}}^{\mathrm {SM}} = 1.
\end{equation}

The third observable $R_b =
 \Gamma_b/\Gamma_{had}$
 has recently received wide
 attention from many quarters and its experimental measurement also
has improved of late quite significantly with the improvement in the
$b$-tagging efficiency. Its expression in the SM
is given by
\begin{equation}
{R_b}^{\mathrm {SM}} \simeq 0.220 + 0.25 ~\Delta V_b^t.
\end{equation}
It may be noted that $R_b$ is free from $\Delta\rho$ and
 $\alpha_S$ (hard bremsstrahlung).

The fourth observable is \cite{girardi}
\begin{equation}
 T = \frac{3}{59} \frac{\Gamma_{had}}{\Gamma_e} -
 \frac{30}{59} \gamma_e.
\end{equation}
Within the SM
\begin{equation}
T^{\mathrm {SM}} \simeq \frac{29}{59} + \frac{19}{59}
 \Delta V_b^t + \frac{\alpha_S}{\pi}.
\end{equation}
In the theoretical expression of $T$, the $\Delta \rho$ term drops
out.

Although $\Delta V_b^t$
contains a quadratic $m_t$ dependence, it
is free from $M_H$ to a significant extent,
and is also reasonably clean from the
kinds of new physics we are looking at.
 On the other hand, $\Delta\rho$,
in addition to having a quadratic $m_t$ dependence,
has a reasonable amount of $M_H$ dependence (logarithmic);
additionally, it
receives tree-level contributions from the types of new physics under
our consideration. So, although the presence of
 $\Delta V_b^t$ induces
some amount of uncertainties through $m_t$, still
 the above $\Delta\rho$-free observables can be
used as `clean microscopes' for looking at physics beyond the SM.

The present stage of experimental precision at LEP
 (after the analysis of the 1992 results) yields:
\begin{eqnarray}
{\widetilde{\gamma}_e}^{\mathrm {exp}} & = & 0.396 \pm 0.004
  \nonumber \\
{R_{\tau e}}^{\mathrm {exp}} & = & 0.996 \pm 0.007 \nonumber \\
{R_b}^{\mathrm {exp}} & = & 0.2200 \pm 0.0027  \\
T^{\mathrm {exp}} & = & 0.513 \pm 0.005. \nonumber
\end{eqnarray}

We now intend to use these `microscopes' to search
 for a new physics
 scenario comprising extra neutral gauge bosons and
  exotic fermions.
We will examine each type of extension {\it one at a time} to
 prevent possible conspiring interplay among the different new
 parameters which will be present in the extended models. Effects
of the simultaneous presence of more than one new physics will also
be discussed in some cases.


\section{$Z^0$--${Z^0}^\prime$ mixing}
\par

Additional gauge bosons besides the
 ones predicted by
the SM are present in many theoretically
 appealing extensions, which project
out the SM as their low-energy manifestation.
 The Grand Unified Theories
(GUTs) are potentially viable candidates
for providing such neutral bosons.
 Many non-GUT models are also there with their
  own justification,
such as the LRS model, which also accommodate
 additional gauge bosons. The additional gauge bososns can
  be neutral
 as well as charged.  For
the present purpose we consider the effects of extra
 ${Z^0}^\prime$ bosons only
as small perturbations to the SM. If, in particular, the gauge group
is $SU(2) \otimes \Pi_{\alpha=1}^n U(1)_\alpha$, there are $n-$ 1
extra ${Z^0}^\prime$ bosons.
 Considering the fact that the couplings between the
extra $U(1)$ bosons and the fermions are quite arbitrary, there are
many unknown parameters, and the phenomenological analysis becomes
cumbersome. But if we take $U(1)$'s as stemming from an underlying
non-Abelian gauge group, in which there is only one overall coupling
constant, the analysis becomes rather simple. The couplings of the
extra ${Z^0}^\prime$ bosons are then predicted from the GUT; the exact
strengths are of course dependent on the symmetry breaking chains.
When the gauge group $SU(2) \otimes U(1)^n$ breaks, at each stage of
symmetry-breaking the corresponding intermediate vector bosons
acquire masses. As a result of the diagonalization of the matrix
consisting of the mass terms of the SM vector bosons, the masses of
the extra gauge bosons and most importantly the off-diagonal terms
that determine the strengths of mixings between the sequential
and additional gauge bosons, the physical gauge boson states are
different from the SM ones.

For the sake of simplicity we will consider
the effects of only {\it one} extra ${Z^0}^\prime$ and that
of only two
origins: (i) $E_6$ breaking down to the SM gauge group via
intermediate breaking steps and yielding extra $U(1)$ group(s)
(extra $U(1)$ models) and (ii) the LRS model with the gauge group
$SU(2)_L\otimes SU(2)_R \otimes U(1)$ breaking in one step to the SM.
In addition to taking the SM one-loop effects, we consider only the
tree-level mixing effects of the additional ${Z^0}^\prime$ with the SM
$Z^0$ without making any assumption of the underlying Higgs structure
responsible for the symmetry-breaking pattern.

The mass eigenstates ($Z, Z^\prime$) are obtained from the gauge
eigenstates ($Z^0, {Z^0}^\prime$) by making the following rotation
\begin{equation}
\pmatrix{~Z \cr ~Z^\prime}= \pmatrix {~\cos \xi_0 & \sin \xi_0~
\cr ~-\sin \xi_0 & \cos \xi_0~}~\pmatrix
{~Z^0 \cr {Z^0}^\prime~}
\end{equation}

Apart from the introduction of this new mixing angle $\xi_0$,
 the mass of
 the physical $Z$ is changed from that of the SM
  $Z^0$. The latter effect is
 manifested through a change of the tree-level $\rho$ parameter.
 The modified vector
 ($v_i$) and axial vector ($a_i$) couplings of
  the physical $Z$ to the
standard fermions are given by
\begin{eqnarray}
v_i & = & \sqrt{\rho} \left[t_3^i - 2 Q_i \STEE +
\xi_0~g^\prime_{vi}\right]
   \nonumber \\
a_i & = & {\sqrt\rho} \left[t_3^i + \xi_0~g^\prime_{ai}\right]
\end{eqnarray}
where $\rho = 1 + \Delta\rho = 1 + \Delta\rho_{\mathrm {tree}} +
\Delta\rho_t = 1+ \Delta\rho_{\mathrm {SB}} +
 \Delta\rho_{\mathrm {M}}
+ \Delta\rho_t$ and $\STEE \simeq
 S_0^2 - \frac{3}{8} \Delta\rho$.
 The tree-order symmetry breaking (SB) and mixing (M)
  contributions, $\Delta \rho_{\mathrm {SB}}$ and $\Delta
\rho_{\mathrm {M}}$ respectively,
  need not be evaluated separately, since
 $\Delta\rho$ does not appear explicitly in the expressions
for the four observables chosen in this study.
We examine only the mixing angles, that
 cast observable consequences on those observables.

For the extra $U(1)$ model
\begin {eqnarray}
g^\prime _{vu} & = & 0 \nonumber \\
g^\prime _{au} & = & {2\over 3} \SW \CT \nonumber \\
g^\prime _{vd} & = & {1\over 2} \SW \left(\CT +\sqrt{5\over 3}
                    \ST\right) \nonumber \\
g^\prime _{ad} & = & -\SW \left({-1\over 6}\CT + {1\over
                     2}\sqrt{5\over 3} \ST \right) \\
g^\prime _{ve} & = & -{1\over 2} \SW \left(\CT + \sqrt{5\over 3}
                     \ST \right) \nonumber \\
g^\prime _{ae} & = & g^\prime _{ad} \nonumber \\
g^\prime _{v\nu} & = & -\SW \left({1\over 6}\CT + {1\over
                     2}\sqrt{5\over 3} \ST \right) \nonumber \\
g^\prime _{a\nu} & = &  g^\prime _{v\nu} \nonumber
\label{gpu1}
\end {eqnarray}
where $\theta_2$ is an angle characteristic of a given
$E_6$-symmetry-breaking chain yielding an extra $U(1)$. Four such
models are of a general phenomenological
interest corresponding to $\theta_2 =
 0^\circ, 52.24^\circ, -52.24^\circ$
and $-82.76^\circ$.

In the LRS model, with $\lambda = g_L/g_R$
 and $y = \sqrt{\CTW - \lambda^2 \STW}$,
\begin {eqnarray}
g^\prime _{vi} & = & {\CTW \over \lambda y}\left(t^i_{3R} - 2
\lambda ^2 Q_i \STW\right) + {\lambda \STW \over y} \left(t^i_{3L} -
2Q_i\STW\right) \nonumber \\
g^\prime _{ai} & = & - {\CTW \over \lambda y} t^i_{3R} +
{\lambda \STW \over y} t^i_{3L}.
\label{gplrs}
\end {eqnarray}

For all practical purposes, one can put $\STW = S_0^2 = 0.23$ in
eqs. (17) and (18).


\section{Ordinary-exotic fermion mixing}
\par

When the bigger groups discussed above break
 down to the SM gauge group, in addition to yielding the extra
gauge bosons, they also entail some additional fermions with various
$SU(2) \otimes U(1)$ representations.
We classify all fermions as either `ordinary'
or `exotic' according to their transformation properties under
$SU(2)$. All the standard fermions, i.e. those contained
in the SM, are called `sequential' fermions.  Their left-chiral
fields transform as doublets and right-chiral fields as
singlets under $SU(2)$. In addition, we consider three other kinds of
non-sequential fermions. These are (i) vector singlets, which
transform as singlets under $SU(2)$ both in the left- and
right-handed sectors, (ii) vector doublets, which transform as
doublets under $SU(2)$ both in the left- and right-handed sectors and
(iii) mirror fermions, which transform as singlets in the left-handed
sector and doublets in the right-handed ones.
 It may be noted that fermions
transforming identically in the two sectors (left and right) do not
have any axial couplings with the gauge bosons and are called
`vector' particles. To complete the nomenclature, we define all
left-handed fermions occurring in doublets (irrespective of whether
they are members of sequential or vector doublets) as `ordinary'
and all left-handed singlets (mirror families or vector singlets) as
`exotic'. Similarly we define all right-handed singlets
(sequential or vector singlets) as `ordinary' and all right-handed
doublets (mirror families or vector doublets) to be `exotic'. In
general, one could speculate on larger varieties of {\it exotica}.
But we have constrained ourselves to only the three types mentioned
above, which follow from various theoretical extensions
beyond the SM.  Our analyses are, however, quite general as they
correspond to a wide class of models without relying specifically on
particular ones.  These fermions are quite heavy (at least heavier
than $M_Z/2$): otherwise they would have been produced in pairs at
LEP. So the only way they can manifest
themselves at LEP is through their mixing with their sequential
counterparts having the same electric charge and colour quantum
numbers.  Fermions with exotic charges and colour assignments do not
mix with the standard fermions as $U(1)_{\mathrm {em}}$
 and $SU(3)_c$ are
unbroken.  These types of fermions have, therefore, not been
considered in our analysis.

To simplify our analysis we assume that exotic fermions mix
 with standard ones
{\it diagonally}, i.e. one exotic fermion mixes with a unique
standard flavour.
 This assumption automatically ensures the absence of tree
level flavour-changing neutral currents (FCNCs)
between light fermion generations,
 which have extremely
tight experimental constraints. We also assume, to make life
 simpler, that there
is only one exotic fermion at a time in the theory which mixes
only with the {\it third}-generation flavour eigenstate. The
loop effects of these fermions are also not considered.

In the presence of mixing the neutral current for the quarks and the
 charged leptons at tree level reads,
\begin{equation}
J^\mu_Z = \frac{g}{{\cos} \theta_W} \left[\overline \psi_{iL}
\gamma^\mu \left(t_3^i {C_L^i}^2 - Q_i \STW \right) \psi_{iL}
 + \overline \psi_{iR} \gamma^\mu
\left(t_3^i {S_R^i}^2 - Q_i \STW \right) \psi_{iR}\right],
\label{ngen2}
\end{equation}
where $C_L^i \equiv \cos \theta_L^i$ and $S_R^i \equiv \sin
\theta_R^i$; $\theta_L^i$ and $\theta_R^i$ are interpreted as
light--heavy mixing angles in the left- and right-handed sectors
respectively.
The ${C_L^i}^2$ term represents a non-universal reduction of strength
of the normal neutral current due to mixing with left-handed
singlets, and the ${S_R^i}^2$ term represents an induced right-handed
current, which is generated as a result of mixing with right-handed
doublets.

The effective vector and axial vector couplings of the $Z$ to the
fermion $i$ follow immediately as
\begin{eqnarray}
 v_i &= &\sqrt{\rho}~ \left[t_3^i\left({C_L^i}^2 +
{S_R^i}^2\right) - 2 Q_i \STE\right] \nonumber \\
 a_i & = &\sqrt{\rho}~ t_3^i\left({C_L^i}^2 - {S_R^i}^2\right).
\label{coupl}
\end{eqnarray}

It may be noted that the SM couplings (given in eq. (\ref{smcoupl}))
can be recovered from eq. (\ref{coupl}) by setting $\theta_L =
\theta_R = 0$. Note that such mixing angles have observable
 consequences only when there are mixings between states with
  different $t_3$. For example, when a $b$ quark
 mixes with a singlet $h$ (charge $-1/3$ non-sequential,
  which stems from $E_6$), only $\theta_L$ is
non-trivial while when $b$ mixes with a $B$ quark sitting in a vector
doublet, only $\theta_R$ is relevant.
  Similar arguments hold for the leptonic sector as well.
We do not treat the mirror fermions separately as their mixing
 effects will always mimic the joint impact of mixings
  in the left- and right-handed
 sectors that we consider case by case in this analysis.

 The mixing effects of the neutrinos
follow the same textures as those of the charged fermions. But
the framework is a bit more complicated due to the presence
of the Majorana mass terms of the neutrinos and due to the lack of
experimental constraints on neutrino FCNCs. In the present analysis
we do not deal with neutrino mixings.


\section{Results}
\par

The reduced electronic width can provide
 the cleanest
(because ordinary--exotic fermion mixings are negligibly small in the
first two generations)
and the most severe bounds on the possible $Z^0$--${Z^0}^\prime$
mixing effects.
Such kinds of mixings modify the reduced electronic width as
(using first-order approximation)
\begin{eqnarray}
\widetilde {\gamma}_e &\simeq &
{\widetilde {\gamma}_e}^{\mathrm {SM}} +
 4 \sin\theta_W \left(-\frac{1}{6} \CT + \frac{1}{2}
  \sqrt{\frac{5}{3}}
\ST\right)~ \xi_0 ~~~ (\mbox{\rm extra U(1)} ) \nonumber \\
\widetilde {\gamma}_e & \simeq &
{\widetilde {\gamma}_e}^{\mathrm {SM}} -
2 {\sqrt{\cos 2\theta_W}}~ \xi_0 ~~~(\mbox{\rm LRS}).
\end{eqnarray}
The bounds are displayed in Table 1.
It is seen that for different
models under consideration, $|\xi_0|$ lies in the range of
 (1--5)$\%$ at 95$\%$ confidence level (C.L.).
These bounds are stronger than the ones in the previous
 analyses, owing to
the reduction of the systematic and statistical errors of the LEP
measurements.

If the $\tau$ lepton mixes with a vector singlet (or with its exotic
partner sitting in a vector doublet), then
 neglecting the $Z^0$--${Z^0}^\prime$ mixing, one obtains,
\begin{eqnarray}
\widetilde {\gamma}_\tau & \simeq &
 {\widetilde {\gamma}_e}^{\mathrm {SM}}
- 2 \SLT~~~(\mbox{\rm singlet})  \nonumber \\
\widetilde {\gamma}_\tau & \simeq &
 {\widetilde {\gamma}_e}^{\mathrm {SM}}
- 2 \SRT~~~(\mbox{\rm doublet})
\end{eqnarray}
When the simultaneous presence
of $\tau$--exotic ($\tau$) and $Z^0$--${Z^0}^\prime$ mixing
is considered, the best
observable to put a
bound on the former, keeping it free from the uncertainties of the
latter (employing the generation universality
 of ${Z^0}^\prime$ coupling to
leptons), is $R_{\tau e}$;
 it is given by
\begin{eqnarray}
 R_{\tau e}& \simeq &1 - 2 \SLT ~~~(\mbox{\rm singlet}) \nonumber \\
 R_{\tau e}& \simeq &1 - 2 \SRT ~~~(\mbox{\rm doublet}).
\end{eqnarray}
for mixing of $\tau$ with a vector singlet (vector doublet).
The bounds at 95$\%$ C.L. are shown in Table 2.

When the $b$ quark mixes with a vector
 singlet (or a member of a vector doublet) partner, and assuming
there is no $Z^0$--${Z^0}^\prime$ mixing,
\begin{eqnarray}
R_b &\simeq & R_b^{\mathrm {SM}} - 0.40~ \SLB~~~(\mbox{\rm singlet})
 \nonumber \\
R_b &\simeq & R_b^{\mathrm {SM}} - 0.08~ \SRB~~~(\mbox{\rm doublet}).
\end{eqnarray}
The bounds at 95$\%$ C.L. of the experimental uncertainties
 are shown in Table 2. It may be noted that
the constraints on $\SLB$ and $\SRB$ are significantly stringent
at $m_t = 200$ GeV compared to the ones for $m_t = 100$ GeV,
because the experimental central value of $R_b$
 at present is close
the theoretical prediction for smaller values of $m_t$.
Also to be noted is that the effect of mixing with a doublet
 is 5 times
less sensitive than with a singlet.

If the $Z^0$--${Z^0}^\prime$ mixing (of extra $U(1)$ type, say)
 and $b$--$h$ (say) mixings are
considered simultaneously, then (using first order approximation)
\begin{equation}
R_b  \simeq  R_b^{\mathrm {SM}} - 0.40~ \SLB + \alpha(\theta_2)~
\xi_0
\end{equation}
 where
\begin{equation}
\alpha(\theta_2)  \simeq  \SW \left(-0.28~\CT + 0.04~\ST\right).
\end{equation}
Just to feel the numerical impact of such simultaneous mixing we take,
as an example, the $\theta_2 = 0^\circ$ model, $m_t = 100$ GeV,
 and employ
the bounds on $|\xi_0|$ from $R_{\tau e}$, to
 obtain $\SLB \leq 0.028$ at 95$\%$ C.L.
 As a result the bound is seen to
  have been relaxed when one compares it with the
corresponding one, namely $\SLB \leq 0.011$ (see Table 2),
 in the absence of
$Z^0$--${Z^0}^\prime$ mixing.

The expression of the
$T$ parameter in the presence of mixing of
a $b$ quark with a singlet becomes
\begin{equation}
 T \simeq T^{\mathrm {SM}}  - \frac{30}{59} \SLB.
\end{equation}
The corresponding bounds at 95$\%$ C.L. are shown in Table 2.
We choose $\alpha_S = 0.12$. Uncertainties due to $\alpha_S$ are small.
 Mixing with a vector doublet,
 as has been seen in the context of $ R_b$,
 relaxes the bound by a factor of 5. Thus all the
bounds on ordinary--exotic fermion mixing angles lie in the range of
(1--5)$\%$ at 95$\%$ C.L. except
for mixing with vector doublets, where the mixing angles
could be large.


\section{Conclusion}
\par
To conclude, we have examined the $Z^0$--${Z^0}^\prime$ and
 ordinary--exotic
mixings in the light of the precision data obtained from $\sim$ 5
million $Z$ events.
 These bounds have been derived from variables that do not depend
on $\Delta\rho$.
 Most of these mixing angles, including all $Z^0$--${Z^0}^\prime$ ones,
  are found to be almost vanishing. For these cases it may be argued
 that to keep the mass(es) of $Z^\prime$ or of the corresponding
exotic fermions in the accessible range of the forthcoming colliders,
favourable choices of the underlying Higgs structure are necessary;
these require closer scrutiny. On the other hand, some mixing angles
(particularly the right-handed ones for $b$ quark) could still be
significantly large. Further reduction of the systematic
 and statistical
errors of the LEP measurements and/or the discovery of the top quark
would definitely reheat these issues in attempts to search for exotic
fermions through such indirect probes.

\vskip 10pt
\noindent{\bf Acknowledgements}
\par

I thank Sunanda Banerjee and Amitava Raychaudhuri
for critically reading the manuscript and for
making important suggestions.
This work has been supported in part by the World Laboratory.

\newpage

\begin{table}[htbp]
\begin{center}
\caption[] {Upper bounds on $Z^0$--${Z^0}^\prime$ mixing angles at
95$\%$ C.L. The observable
 from which the bounds are derived is $\widetilde{\gamma}_e$.}
\bigskip
\begin{tabular}{|c|c|c|c|c|c|}
\hline
\multicolumn{5}{|c|}{Extra $U(1)$} & LRS \\
\hline
$\theta_2$ (deg) & 0 & 52.24 & $-52.24$ & $-82.76$ & $\lambda = 1$ \\
\hline
$|\xi_0|$ (rad) & 0.05 & 0.02 & 0.02 & 0.01 & 0.01 \\
\hline
\end{tabular}
\end{center}
\end{table}

\begin{table}[htbp]
\begin{center}
\caption[] {Upper bounds on ordinary--exotic mixing angles
 at 95$\%$ C.L.
  The observables
 from which the bounds are derived are mentioned. For
$b$ quark mixing angles the bounds refer to $m_t = 100~(200)$ GeV,
and they correspond to the situation when there is no
$Z^0$--${Z^0}^\prime$ mixing.}
\bigskip
\begin{tabular}{|c|c|c|c|c|c|}
\hline
\multicolumn{2}{|c|}{$R_{\tau e}$} &
\multicolumn{2}{c|}{$R_b$} & \multicolumn{2}{c|}{$T$} \\
\hline
$\SLT$ & $\SRT$ & $\SLB$ & $\SRB$  & $\SLB$ & $\SRB$ \\
\hline
0.009 & 0.009  & 0.011 & 0.055 & 0.05 & 0.25 \\
      &        & (0.001) & (0.005) & (0.04) & (0.20) \\
\hline
\end{tabular}
\end{center}
\end{table}

\end{document}